\documentstyle[12pt]{article}

\begin{document}

\begin{flushright}
GUTPA/01/03/01\\
\end{flushright}
\vskip .1in

\begin{center}
{\Large {\bf Spontaneously Generated Gauge Invariance}}

\vspace{35pt}

{\bf J.L. Chkareuli}

\vspace{6pt}

{\em Institute of Physics, Georgian Academy of Sciences, 380077 Tbilisi,
Georgia\\[0pt]
}

\vspace{12pt}

{\bf C.D. Froggatt}

\vspace{6pt}

{\em Department of Physics and Astronomy\\[0pt]
Glasgow University, Glasgow G12 8QQ, Scotland\\[0pt]
}

\vspace{12pt}

{\bf H.B. Nielsen} 
\vspace{6pt}

{\em Niels Bohr Institute, \\[0pt]
Blegdamsvej 17-21, DK 2100 Copenhagen, Denmark}

\bigskip

{\large {\bf Abstract}}
\end{center}

We argue that the non-observability of the spontaneous breakdown of Lorentz
symmetry (SBLS) caused by the vacuum expectation values of vector fields
could provide the origin of all internal symmetries observed. Remarkably,
the application of this principle to the most general relativistically
invariant Lagrangian, with arbitrary couplings for all the fields involved,
leads by itself to the appearance of a symmetry and, what is more, to the
massless vector field(s) gauging this symmetry. A simple model for the SBLS
based on massive vector and real scalar field interactions is considered;
it is found that spontaneously broken gauge symmetries could also
appear, when SBLS happens and is required to be physically unobservable.

\thispagestyle{empty} \newpage

\section{Introduction \label{Introduction}}

There are several indications in the literature (see \cite{book} and
extended references therein) that a local symmetry of all fundamental
interactions of the matter and the corresponding massless gauge fields could
be dynamically generated. They include observations stemming from
Kaluza-Klein type theories, string theories, non-linear $\sigma $-models and
non-standard interpretations of gauge symmetry.

In particular there has been considerable interest \cite{bj} in the
interpretation of gauge fields as composite Nambu-Jona-Lasinio (NJL) bosons
\cite{njl}, possibly associated with the spontaneous breakdown of Lorentz
symmetry (SBLS). A typical model realising this mechanism is based on the
four fermion (current $\times $ current) interaction, where the gauge field
appears as a fermion-antifermion pair composite state, in complete analogy
with the massless composite scalar field in the original NJL model
\cite{njl}.
While this model is non-renormalisable and one must fix the momentum
cut-off at some large (but finite) scale $\Lambda $, it is well-known to be
largely equivalent to a gauge theory under the compositeness condition $%
Z_{3}=0$, where $Z_{3}$ is the wave function renormalisation constant of the
gauge boson \cite{eguchi}. Under this condition, the gauge field becomes an
auxiliary field without propagating degrees of freedom. The quantum
fluctuations, then, give rise to its kinetic and self-interaction (for the
non-Abelian case) terms, so that a dynamical gauge boson is finally induced
as an independent composite field. However, in contrast to the belief
advocated in the pioneering works \cite{bj}, there appears a generic problem
in turning the composite vector particles into genuine massless gauge bosons
\cite{suzuki}. Actually, one must make a precise tuning of parameters,
including a cancellation between terms of different orders in the $1/N$
expansion (where $N$ is the number of fermion species involved), in order to
achieve the massless case needed.

In this note we would like to turn back to the role of Lorentz symmetry in a
dynamical generation of gauge invariance. We argue that, generally, Lorentz
invariance can only be imposed in the sense that all Lorentz non-invariant
effects caused by its spontaneous breakdown are physically unobservable.
We show here that the physical non-observability of the spontaneous
breakdown of Lorentz symmetry (SBLS), taken as a basic principle, leads to
genuine gauge invariant theories in both Abelian (Section 2) and non-Abelian
(Section 3) cases, even though one starts from an arbitrary relativistically
invariant Lagrangian. In the original Lagrangian, the vector fields are
taken as massive and all possible kinetic and interaction terms are
included.
However, when SBLS occurs and its non-observability is imposed, the vector
bosons become massless and the only surviving interaction terms are those
allowed by the corresponding gauge symmetry. Thus, the Lorentz symmetry
breaking does not manifest itself in any physical way, due to the generated
gauge symmetry converting the SBLS into gauge degrees of freedom of the
massless vector bosons. Remarkably, even global symmetries are not required
in the original Lagrangian - the SBLS induces them automatically.

A simple heuristic model for the SBLS, based on massive vector and real
scalar field interactions, is considered (Section 4) and it is found that
spontaneously broken gauge symmetries could also appear when SBLS
occurs and is required to be physically unobservable.

Finally we summarise our conclusions in Section 5.

\section{Abelian gauge invariance \label{abelian}}

The would-be Abelian model in the simplest case corresponds to massive
electrodynamics including one charged fermion species. The Lagrangian
density for the model is given by:
\begin{equation}
L=-\frac{1}{4}F_{\mu \nu }F_{\mu \nu }+\frac{1}{2}M^{2}A_{\mu }^{2}+i%
\overline{\psi }\gamma _{\mu }\partial _{\mu }\psi -m\overline{\psi }\psi
+eA_{\mu }\overline{\psi }\gamma _{\mu }\psi  \label{L}
\end{equation}
where $e$ is the appropriate (would-be gauge) coupling constant of the
vector field $A_{\mu }$ with the charged fermion field.
In this paper we shall assume that all the vector fields describe
pure spin-1 fields satisfying the Lorentz condition:
\begin{equation}
\partial _{\mu }A_{\mu }=0  \label{t}
\end{equation}
We impose this condition (\ref{t}) as an off-shell constraint, singling
out a genuine spin-1 component in the four-vector $A_{\mu }$ independent
of the equations of motion.
One can readily see
that the Lagrangian density (\ref{L}) possesses a $U(1)^{\psi }$ global
symmetry, with the corresponding conserved Noether current $j_{\mu }=
\overline{\psi }\gamma _{\mu }\psi $ which, in this simple case, is
closely connected to
the current $j_{\mu }^{A}$ associated with $A_{\mu }$,

\begin{equation}
j_{\mu }^{A}=\frac{\partial L}{\partial A_{\mu }}=ej_{\mu } + M^2 A_{\mu}
\label{j}
\end{equation}
It is then clear that
the $j_{\mu }^{A}$ current is also conserved, provided that
the vector field satisfies the Lorentz condition (\ref{t})
and preserves its transverality.

Let us consider now the SBLS in some detail. We propose that the vector
field $A_{\mu }$ takes the form
\begin{equation}
A_{\mu }=a_{\mu }(x)+n_{\mu }  \label{f}
\end{equation}
when the SBLS occurs. Here the constant Lorentz four-vector $n_{\mu }$ is a
classical background field appearing when the vector field $A_{\mu }$
develops a VEV. We do not yet specify the mechanism which could induce the
SBLS of type (\ref{f})---rather we study its general consequences for the
possible dynamics of the matter and vector (gauge) fields and require
it to be physically unobservable.

Substitution of the form (\ref{f}) into the Lagrangian (\ref{L}) immediately
shows that the kinetic term for the vector field $A_{\mu }$ translates into
a kinetic term for $a_{\mu }$ ($F_{\mu \nu }^{(A)}=F_{\mu \nu }^{(a)}$),
while its mass and interaction terms are correspondingly changed:
\begin{equation}
L=-\frac{1}{4}F_{\mu \nu }F_{\mu \nu }+\frac{1}{2}M^{2}(a_{\mu }+n_{\mu
})^{2}+i\overline{\psi }\gamma _{\mu }\partial _{\mu }\psi -m\overline{\psi
}%
\psi +e(a_{\mu }+n_{\mu })\cdot \overline{\psi }\gamma _{\mu }\psi
\label{L*}
\end{equation}
As to the interaction term, one can always go by a unitary
transformation\footnote{In momentum representation this
transformation corresponds to displacing the momentum of each
fermion by an amount $en_{\mu}$. In the presence of terms breaking
the $U(1)^{\psi}$ global symmetry, this transformation would
induce a breaking of momentum conservation in processes where
the $U(1)^{\psi}$ charge is not conserved. This non-conservation of
momentum would also break Lorentz symmetry, whose breaking we require to be
unobservable, and so it is necessary here to have $U(1)^{\psi}$ charge
conservation.\label{foot1}}
to a new fermion field ${\bf \Psi }$

\begin{equation}
\psi = \exp [ie\omega (x)]\textrm{ }{\bf \Psi }\textrm{, \qquad }
\omega(x)=n\cdot x  \label{psi}
\end{equation}
so as to exactly cancel the Lorentz symmetry-breaking term $n_{\mu }\cdot
\overline{\psi }\gamma _{\mu }\psi $ in the Lagrangian (\ref{L*}). This
cancellation occurs due to the appearance of a compensating term from the
fermion kinetic term, provided that the phase function $\omega (x)$ is
chosen to be linear in the coordinate four-vector $x_{\mu }$ (as indicated
in Eq.~\ref{psi})\footnote{A gauge function of this type
was first considered in the framework of
quantum electrodynamics \cite{ferrari}.
We use this form for $\omega $ here
just for simplicity and convenience. In the general case, when the
shift in the vector field depends on space-time in the form
$A_{\mu }=a_{\mu }(x)+n_{\mu }f(n\cdot x)$ where $f$ is
any integrable function of $y=n\cdot x=n_{\mu }x_{\mu }$, the gauge
function $\omega $ is given by the integral $\omega =\int f(y)dy$. In
the non-Abelian case (see Section 3) such a shift could have the general
factorised form $A_{\mu }^{i}=a_{\mu }^{i}(x)+n_{\mu }f^{i}(n\cdot x)$,
with the gauge function $\omega ^{i}$ given by
the integral $\omega ^{i}=\int f^{i}(y)dy$. The corresponding vacuum
states would of course formally break translational invariance, as
well as Lorentz invariance.}.
Thus, the only trace left of SBLS
is in the mass term for the vector field which contains the scalar product
$(n\cdot a)$ and the physical non-observability of SBLS requires
it to be set equal to zero:

\begin{equation}
M^{2}(n\cdot a)=0  \label{con}
\end{equation}
An extra gauge condition $n\cdot a$ $\equiv n_{\mu }\cdot a_{\mu }=0$ would
be incompatible with the Lorentz gauge (\ref{t}), which has already been
imposed for the massive vector field $a_{\mu }$. Therefore, the only way to
satisfy Eq.~(\ref{con}) is to take $M^{2}=0$. Otherwise Lorentz symmetry is
explicitly broken. In such a way all other terms but the gauge invariant
ones (for which ``compensating'' local transformations of the type
(\ref{psi}) are always available) can be excluded by our
non-observability assumption.

Let, for example, the original Lagrangian include a term of the type

\begin{equation}
\Delta L_{1}=\frac{f}{4}A_{\mu }^{2}\cdot A_{\mu }^{2}  \label{AAAA}
\end{equation}
which would cause non-conservation of the current

\begin{equation}
j_{\mu }^{A}=ej_{\mu }+M^2A_{\mu}+fA_{\mu }\cdot A_{\nu }^{2}
\label{current}
\end{equation}
related with the vector field $A_{\mu }$. The condition (\ref{con}) for the
non-observability of SBLS now becomes

\begin{equation}
\lbrack M^{2}\textrm{ }+f(a^{2}+(n\cdot a)+n^{2})](n\cdot a)=0  \label{cons**}
\end{equation}
Again, this condition must be fulfilled identically, i.e.~$M^{2}=0$ and
$f=0$. Otherwise it would either represent
another supplementary condition on $a_{\mu}$, in addition to
the Lorentz gauge already taken (\ref{t}), or it would impose another
dynamical equation in addition to the usual Euler equation for the vector
field $a_{\mu }$. Hence physical Lorentz invariance requires the survival of
$j_{\mu }^{A}$ current conservation.

So far we have considered (a) the case of massive electrodynamics where the
current $j_{\mu }^{A}$ related to the vector field is conserved, and (b) the
case where $j_{\mu }^{A}$ would not be conserved due to the vector field
self-interaction term (\ref{AAAA}), while the Noether current $j_{\mu }=%
\overline{\psi }\gamma _{\mu }\psi $ is still conserved. Let us now include
in the Lagrangian (\ref{L}) terms which break even global $U(1)^{\psi }$
invariance:

\begin{equation}
\Delta L_{2}=\frac{G}{2}A_{\mu }\overline{\psi }_{C}\gamma _{\mu }\gamma
_{5}\psi +\frac{G^{*}}{2}A_{\mu }\overline{\psi }\gamma _{\mu }\gamma
_{5}\psi _{C}  \label{CC}
\end{equation}
where $\psi _{C}$ is a charge-conjugated spinor, $\psi _{C}=C\overline{\psi
}$, while $G$ and its complex conjugate $G^{*}$ stand for coupling constants
(the term $A_{\mu }\overline{\psi }_{C}\gamma _{\mu }\psi $ and its
Hermitian conjugate are absent since they are identically equal to zero due
to Fermi statistics). So, in writing down the Lagrangian (\ref{L}) with the
addition of $\Delta L_{2}$, we do not impose beforehand any restrictions
connected with conservation of fermion number. Now, in distinct contrast to
the ordinary vector field-current interaction in Lagrangian (\ref{L}), no
transformation of the type (\ref{psi}) is available for the fermion field $%
\psi $ which could eliminate the trace of SBLS in $\Delta L_{2}$ (\ref{CC}).
The Lorentz invariance condition (\ref{cons**}), extended now so as to
include the SBLS terms from $\Delta L_{2}$, again requires the identical
vanishing of the vector boson mass $M^{2}=0$ and coupling constants, $f=0$
and $G=0$. Otherwise there would be fewer degrees of freedom for the vector
and fermion fields, $a_{\mu }$ and $\psi $, than is needed for describing
the spins $1$ and $1/2$, respectively, which is
inadmissible\footnote{If we extend the SBLS non-observability assumption
to the much stronger form required in the next section to derive a
non-Abelian gauge symmetry, it is not necessary to impose the
the Lorentz condition (\ref{t}) as an
off-shell constraint on the four-vector field $A_{\mu }$.
The stronger form of the SBLS non-observability
assumption requires the condition (\ref{con}) or (\ref{cons**}) to
be satisfied for {\em any} vector $n_{\mu }$. In this case, the
condition $n\cdot a$ $\equiv n_{\mu }\cdot a_{\mu }=0$ would
require the vector field to vanish identically, $a_{\mu }=0$,
which is also inadmissable. When the Lorentz condition (\ref{t})
is dropped, there is an extra allowed contribution
$\beta \cdot \left( \partial_{\mu }A_{\mu }\right) ^{2}$
to the kinetic term for the vector field; however the SBLS
non-observability assumption would require it to vanish
($\beta =0$), if the vacuum vector field took the general
space-time dependent form considered in footnote 2.}.

Proceeding in such a way with all possible interactions (including those
with other fermion and scalar fields), one finally arrives at the gauge
invariant Abelian theory as the only version of the theory which is
compatible with physical Lorentz invariance when SBLS occurs.
One may, of course, wonder whether it is at all possible for the
lowest energy (vacuum) state to have SBLS. It is indeed
possible, by introducing an $\frac{f}{4}A_{\mu}^2\cdot A_{\mu}^2$ term
(\ref{AAAA}), to arrange that the $A_{\mu}$ field
potential energy density term
is minimised for $n_{\mu}^2 = -\frac{M^2}{f}$. However we have seen that
the non-observability of SBLS requires $M^2 = f = 0$, in which case
the (vanishing) vector field potential obviously has many flat directions
and the energy density for the SBLS and Lorentz invariant vacua
happen to be the same. This means that the SBLS vacuum is quite possible.

We will extend this discussion to include scalar fields and the
spontaneous breakdown of the gauge symmetry in section \ref{scalar}.

\section{Non-Abelian gauge invariance \label{nonabelian}}

Let us come now to the many-vector field case which results in the
non-Abelian gauge symmetry. We suppose there are a number of pure spin-1
vector fields of the type $A_{\mu }(x)$, satisfying the Lorentz gauge
condition (\ref{t}), considered in the previous section. Proposing not even
a global symmetry at the start, one can simply collect them in some set $%
A_{\mu }^{i}(x)$ with $i=1,...N$. The matter fields, say fermions again as
in the Abelian case, are collected in another set $\psi =(\psi
^{(1)},...,\psi ^{(r)})$. The general Lagrangian describing all their
interactions with dimensionless coupling constants (some exotic terms
violating fermion number and parity are omitted for simplicity)
is given by:

\begin{eqnarray}
L &=&-\frac{1}{4}F_{\mu \nu }^{i}F_{\mu \nu }^{i}+\frac{1}{2}%
(M^{2})_{ij}A_{\mu }^{i}A_{\mu }^{j}+\alpha ^{ijk}\partial _{\nu }A_{\mu
}^{i}\cdot A_{\mu }^{j}A_{\nu }^{k}+\beta ^{ijkl}A_{\mu }^{i}A_{\nu
}^{j}A_{\mu }^{k}A_{\nu }^{l}+  \nonumber \\
&&{\bf +}i\overline{\psi }\gamma \partial \psi -\overline{\psi }m\psi
+A_{\mu }^{i}\overline{\psi }\gamma _{\mu }T^{i}\psi  \label{LN}
\end{eqnarray}
Here $F_{\mu \nu }^{i}=\partial _{\mu }A_{\nu }^{i}-\partial _{\nu }A_{\mu
}^{i}$ , 
while $(M^{2})_{ij}$ is a general $N\times N$ mass-matrix for the vector
fields and $\alpha ^{ijk}$ and $\beta ^{ijkl}$ are their coupling
constants---all as yet unknown numbers. The $r\times r$ matrices $m$ and $%
T^{i}$ contain, in a compact form, the still arbitrary fermion masses and
coupling constants describing the interaction between the fermions and the
vector fields (all the numbers mentioned are real and the matrices
Hermitian, as follows in this case from the Hermiticity of the Lagrangian).

We assume now that the vector fields $A_{\mu }^{i}$ each take the form
\begin{equation}
A_{\mu }^{i}(x)=a_{\mu }^{i}(x)+n_{\mu }^{i}  \label{ab}
\end{equation}
when SBLS occurs; here the independent constant Lorentz four-vectors $n_{\mu
}^{i}$ ($i=1,...N$) are the VEVs of the vector fields,
analogous to the $n_{\mu }$ of the Abelian case.
However, in contrast to the Abelian case
where the constant vector $n_{\mu }$ in the form (\ref{f}) may have any
length, we consider here at first
just infinitesimally small $n_{\mu }^{i}$ four
vectors (for the generalization to finite vectors $n_{\mu }^{i}$ see below).
Furthermore we require the non-observability of the SBLS for {\em any}
set\footnote{As will be motivated in the conclusion, we are here using the
very strong assumption that the SBLS shall be physically unobservable
whatever the infinitesimal VEVs $n_{\mu}^i$ would be. This means
we are assuming that Nature hides the SBLS not only for the actually
realised vacuum but also for the a priori possible vacua.}
of infinitesimal vectors $n_{\mu }^{i}$.

Substitution of the form (\ref{ab}) into the Lagrangian (\ref{LN}) shows,
again as in the Abelian case, that the kinetic term for
the vector fields $A_{\mu }^{i}$ translates into
a kinetic term for the vector fields $a_{\mu}^{i}$
($F_{\mu \nu }^{(A)}=F_{\mu \nu }^{(a)}$), while their mass and
interaction terms are correspondingly changed:

\begin{eqnarray}
L &=&-\frac{1}{4}F_{\mu \nu }^{i}F_{\mu \nu }^{i}+\frac{1}{2}%
(M^{2})_{ij}(a_{\mu }^{i}+n_{\mu }^{i})(a_{\mu }^{j}+n_{\mu }^{j})
\nonumber
\\
&&+\alpha ^{ijk}\partial _{\nu }a_{\mu }^{i}\cdot (a_{\mu }^{j}+n_{\mu
}^{j})(a_{\nu }^{k}+n_{\nu }^{k}){\bf +}  \nonumber \\
&&+\beta ^{ijkl}(a_{\mu }^{i}+n_{\mu }^{i})(a_{\nu }^{j}+n_{\nu
}^{j})(a_{\mu }^{k}+n_{\mu }^{k})(a_{\nu }^{l}+n_{\nu }^{l})+  \label{lagr}
\\
&&+i\overline{\psi }\gamma \partial \psi -\overline{\psi }m\psi +(a_{\mu
}^{i}+n_{\mu }^{i})\overline{\psi }\gamma _{\mu }T^{i}\psi  \nonumber
\end{eqnarray}
The non-observability of the SBLS for {\em any} set of infinitesimal vectors
$n_{\mu }^{i}$ requires, as in the Abelian case, exact cancellations
between
non-Lorentz invariant terms of the same structure in the Lagrangian (\ref
{lagr}). One can first introduce a new set of
vector fields ${\bf a}_{\mu }^{i}$
defined by the infinitesimal transformation

\begin{equation}
a_{\mu }^{i}={\bf a}_{\mu }^{i}-\alpha ^{ijk}\omega ^{j}(x){\bf a}_{\mu}^{k}
\textrm{ , \quad }\omega ^{i}(x)=n_{\mu }^{i}\cdot x_{\mu }  \label{rot}
\end{equation}
which includes the above coupling constants $\alpha ^{ijk}$ and the linear
``gauge'' functions $\omega ^{i}(x)$. The condition that the Lorentz
symmetry-breaking terms in the trilinear and quadrilinear self-interaction
couplings of the vector fields ${\bf a}_{\mu }^{i}$, including those which
arise from the kinetic term for the vector fields, should cancel for {\em
any} infinitesimal vector $n_{\mu }^{i}$
is then satisfied, if and only if the
sets of real numbers $\alpha ^{ijk}${\bf \ }and $\beta ^{ijkl}$ in these
couplings satisfy the following conditions ({\bf a}) and ({\bf b}):

({\bf a}) $\alpha ^{ijk}$ is totally antisymmetric (in the indices $i$, $j$
and $k)$ and obeys the structure relations:

\begin{equation}
\alpha ^{ijk}{\bf \equiv }\alpha ^{[ijk]}\equiv \alpha _{[jk]}^{i}\textrm{ ,
\quad }[\alpha ^{i},\alpha ^{j}]=-\alpha ^{ijk}\alpha ^{k}  \label{alg}
\end{equation}
where the $\alpha ^{i}$ are defined as matrices with elements $(\alpha
^{i})^{jk}=\alpha ^{ijk}$. This means that the matrices $\alpha ^{k}$ form
the adjoint representation of a Lie algebra, under which the vector fields
transform infinitesimally as given in Eq.~(\ref{rot}). In the case when the
matrices $\alpha ^{i}$ can be decomposed into a block diagonal form, there
appears a product of symmetry groups rather than a single simple
group\footnote{For simplicity we shall consider the
case of a single simple group in the following.}.

({\bf b}) $\beta ^{ijkl}$ takes the factorised form

\begin{equation}
\beta ^{ijkl}= - \frac{1}{4} \alpha ^{ijm}\cdot \alpha ^{klm}  \label{fac}
\end{equation}

The above requirements ({\bf a}) and ({\bf b}) would of course also be
fulfilled if the coupling terms $\alpha_{ijk}$ were zero rather than
antisymmetric structure constants, as is required to make the theory a true
Yang-Mills theory. We would however like to argue that it is most natural
and most likely that the gauge theory derived will be truly non-Abelian.
Really it should be sufficient to just remark that there are many more
possibilities for interacting Yang-Mills theories than for vector fields
without self-interactions.

Let us turn now to the mass term for the vector fields
in the Lagrangian (\ref{lagr}). When expressed in terms
of the transformed vector fields ${\bf a}_{\mu }^{i}$
(\ref{rot}), it contains should-be
vanishing SBLS remnants of the type

\begin{equation}
(M^{2})_{ij}(\alpha ^{ikl}\omega ^{k}{\bf a}_{\mu }^{l}{\bf a}_{\mu }^{j}%
\textrm{ }{\bf +a}_{\mu }^{i}n_{\mu }^{j}){\bf =}0  \label{mmm}
\end{equation}
Here we have used the symmetry feature $(M^{2})_{ij}=(M^{2})_{ji}$ for a
real Hermitian matrix $M^{2}$ and have retained only the first-order terms
in $n_{\mu}^{i} $. These two types of remnant have different structures and
hence must vanish independently. One can readily see that, in view of the
antisymmetry of the structure constants, the first term in Eq.~(\ref{mmm})
may be written in the following form containing the commutator of the
matrices $M^{2}$ and $\alpha ^{k}$

\begin{equation}
\lbrack M^{2},\alpha ^{k}{\bf ]}_{jl}\omega ^{k}{\bf a}_{\mu }^{l}
{\bf a}_{\mu }^{j}=0  \label{com}
\end{equation}
It follows that the mass matrix $M^{2}$ should commute with all the matrices
$\alpha ^{k}$, in order to satisfy Eq.~(\ref{com}) for all sets of ``gauge''
functions $\omega ^{i}=n_{\mu }^{i}\cdot x_{\mu }$. That means according to
Schur's lemma that, since the matrices $\alpha ^{k}$
heve been shown to form an
irreducible representation of a (simple) Lie algebra, the matrix $M^{2}$ is a
multiple of the identity matrix, ($M^{2})_{ij}={\bf M}^{2}\delta _{ij}$,
thus giving the same mass for all the vector fields. It then follows
that the vanishing of the second term in Eq.~(\ref{mmm}) leads to the simple
condition analogous to that (\ref{con}) in the Abelian case:

\begin{equation}
{\bf M}^{2}(n^{i}\cdot {\bf a}^{i})=0  \label{m}
\end{equation}
for any infinitesimal $n_{\mu }^{i}$. Since the Lorentz gauge condition has
already been imposed on ${\bf a}_{\mu }^{i}$ ($\partial _{\mu }{\bf a}_{\mu
}^{i}=0$), we cannot impose extra gauge conditions of the type $n^{i}\cdot
{\bf a}^{i}=n_{\mu }^{i}\cdot {\bf a}_{\mu }^{i}=0$. Thus, we are
necessarily led to:

({\bf c}) masslessness of the vector fields,
$(M^{2})_{ij}={\bf M}^{2}\delta_{ij\textrm{ }}=0$.

Finally we consider the interaction term between the vector and fermion
fields in the Lagrangian (\ref{lagr}). In terms of the transformed vector
fields ${\bf a}_{\mu }^{i}$ (\ref{rot}), it takes the form

\begin{equation}
({\bf a}_{\mu }^{i}-\alpha ^{ijk}\omega ^{j}{\bf a}_{\mu }^{k}\textrm{ }
+n_{\mu }^{i}{\bf )\cdot }\overline{\psi }\gamma _{\mu }T^{i}\psi
\label{fer}
\end{equation}
One now readily confirms that the Lorentz symmetry-breaking terms (the
second and third ones) can be eliminated, when one goes to a new set of
fermion fields ${\bf \Psi }$ using a unitary transformation of the type:

\begin{equation}
\psi =\exp [iT^{i}\omega ^{i}(x)]{\bf \Psi }\textrm{ , \qquad }\omega
^{i}(x)=n^{i}\cdot x  \label{ff}
\end{equation}
A compensating term appears from the fermion kinetic term and the
compensation occurs for any set of ``gauge'' functions $\omega^i(x)$ if and
only if:

({\bf d}) the matrices $T^{i}$ form a representation of the Lie algebra with
structure constants $\alpha ^{ijk}$

\begin{equation}
\lbrack T^{i},T^{j}]=i\alpha ^{ijk}T^{k}  \label{TTT}
\end{equation}
In general this will be a reducible representation but, for simplicity, we
shall take it to be irreducible here. This means
that the matter fermions ${\bf \Psi }$ are all
assigned to an irreducible multiplet determined by the
matrices $T^{i}$. At the same time, the unitary transformation (\ref{ff})
changes the mass term for the fermions to

\begin{equation}
\overline{{\bf \Psi }}(m+i\omega ^{k}[m,T^{k}]){\bf \Psi }  \label{k}
\end{equation}
The vanishing of the Lorentz non-invariant term (the second one) in
Eq.~(\ref{k}) for any set of ``gauge''
functions $\omega^i(x)$ requires that the
matrix $m$ should commute with all the matrices $T^{k}$. According to
Schur's lemma, this again means that the matrix $m$ is proportional to the
identity, $m_{rs}={\bf m}\delta _{rs}$, thus giving:

({\bf e}) the same mass for all the matter fermion fields within the
irreducible multiplet determined by the matrices $T^{i}$; in the case when
the fermions are decomposed into several irreducible multiplets, their
masses are necessarily equal only within each multiplet.

The above argument for gauge symmetry as a consequence of the
non-observability of SBLS also works in the absence
of matter fields. However, if fermions are present, there is
a technically shorter derivation of all the above conclusions.
Firstly, one can transform to the new sets of vector and fermion
fields according to Eqs.~(\ref{rot}) and (\ref{ff}) and,
as in the Abelian case, consider the
vector field and fermion-current interaction terms (\ref{fer}).
After compensation of all the
linear (under $n_{\mu }^{i}$) SBLS terms, one is led to the commutator (\ref
{TTT}) for the fermion current matrices $T^{i}$. Then the standard Jacobi
identity for the $T$ matrices immediately leads to the basic structure
relation (\ref{alg}) for the $\alpha ^{ijk}$, whose antisymmetry with respect
to indices $i$ and $j$ follows from the commutator (\ref{TTT}) by itself.
Its antisymmetry under indices $j$ and $k$ stems from the
required cancellation of the linear SBLS terms (containing
a vector-field derivative) in the trilinear self-interaction
couplings  in the Lagrangian (\ref{lagr})
with the terms that appear from
the kinetic term for the vector fields. The cancellation of the
linear SBLS terms (containing no vector-field derivative) in the
trilinear and quadrilinear vector field couplings in the
Lagrangian results in the relation (\ref{fac}) for their
coupling constants. Finally, the conclusions ({\bf c}) and ({\bf e}) for the
masses of the vector and fermion fields are derived.

Now, collecting together all the outcomes ({\bf a})-({\bf e}) derived from
the non-observability of the SBLS for {\em any} set of infinitesimal vectors
$n_{\mu }^{i}$ applied to the general Lagrangian (\ref{LN}), we arrive at a
truly gauge invariant Yang-Mills theory for the new fields
${\bf a}_{\mu}^{i}$ and ${\bf \Psi }$:

\begin{equation}
L_{YM}={\bf -}\frac{1}{4}{\bf F}_{\mu \nu }^{i}{\bf F}_{\mu \nu }^{i}+i
\overline{{\bf \Psi }}\gamma \partial {\bf \Psi -m}\overline{{\bf \Psi }}
{\bf \Psi +ga}_{\mu }^{i}\overline{{\bf \Psi }}{\bf \gamma }_{\mu }
{\bf T}^{i}{\bf \Psi }  \label{fin}
\end{equation}
Here ${\bf F}_{\mu \nu }^{i}=\partial_\mu {\bf a}_{\nu }^{i}
{\bf -}\partial_{\nu }{\bf a}_{\mu }^{i}
{\bf +ga }^{ijk}{\bf a}_{\mu }^{j}{\bf a}_{\nu}^{k}$ and ${\bf g}$ is a
universal gauge coupling constant extracted from
the corresponding matrices $\alpha ^{ijk}={\bf ga }^{ijk}$ and
$T^{i}={\bf gT}^{i}$.

Let us now consider the generalisation of the vector field VEVs from
infinitesimal to finite background classical fields $n_{\mu}^i$.
Unfortunately one cannot directly generalise the SBLS form (\ref{ab})
to all finite $n_{\mu }^{i}$ vectors. Due to the non-cancellation of the
high-order $n_{\mu }^{i}$ terms, one would inevitably come to
an Abelian rather than a non-Abelian symmetry, when applying the
principle of the SBLS non-observability to the Lagrangian (\ref{lagr}).
Otherwise, one has a non-vanishing field strength $F_{\mu \nu }^{a}$ in the
vacuum, implying a real physical breakdown of Lorentz symmetry.
This problem can be automatically
avoided if the finite SBLS shift vector $n_{\mu }^{i}$ in the basic equation
(\ref{ab}) takes the factorised form

\begin{equation}
A_{\mu }^{i}(x)=a_{\mu }^{i}(x)+n_{\mu }\cdot f^{i}  \label{fff}
\end{equation}
where $n_{\mu }$ is a constant Lorentz vector as in the Abelian case,
while $f^{i}$ ($i=1,2,...N$) is a vector in the internal charge space.
So we now formulate the strong form of our SBLS non-observability assumption,
needed to derive non-Abelian gauge invariance, as follows:
the SBLS must remain hidden for any set of VEVs for the vector
fields $A_{\mu}^i$ of the factorised form
$n_{\mu }^{i}=n_{\mu }\cdot f^{i}$ (\ref{fff}).
Using the Lagrangian (\ref{fin}) derived for infinitesimal VEVs,
it is now straightforward to show that there will be no observable effects
of SBLS for {\em any} set of finite factorised VEVs
$n_{\mu }^{i}=n_{\mu }\cdot f^{i}$ (\ref{fff}). For this purpose,
it is sufficient to generalise Eq.~(\ref{rot})
to the finite transformation:

\begin{equation}
a_{\mu }\cdot \alpha =\exp [(\omega \cdot \alpha )]{\bf a}_{\mu }\cdot
\alpha \exp [-(\omega \cdot \alpha )]  \label{finiterot}
\end{equation}
Then Eqs. (\ref{ab}, \ref{ff}, \ref{finiterot}) combine to form a genuine
finite gauge transformation\footnote{Note that
$\exp [(n\cdot x)(f\cdot \alpha )]\partial _{\mu }\exp [-(n\cdot
x)(f\cdot \alpha )]=-n_{\mu }(f\cdot \alpha )$ and hence Eq.~(\ref{fff})
correctly represents the derivative term in the gauge transformation for
finite factorised VEVs $n_{\mu}^{i}=n_{\mu}\cdot f^{i}$.} for the Yang-Mills
Lagrangian (\ref{fin}) and the SBLS is simply transformed away as a gauge
degree of freedom.

If other fermion and scalar fields are included, the Lorentz invariance
condition applied to their SBLS remnants will assign them to appropriate
symmetry multiplets. Thus the gauge symmetry is readily extended to the
Yukawa and Higgs sectors as well.

\section{Spontaneously broken gauge invariance \label{scalar}}

So far our considerations have been quite general and model-independent, as
they concerned the impact of the SBLS upon the possible dynamics of the
matter and vector (gauge) fields. However, we have not yet explicitly
discussed the role of scalar fields, particularly in the situation when
these fields develop VEVs. An interesting question then arises: how does the
SBLS interplay with the spontaneously broken internal symmetries?
The presence of scalar fields may also provide a
possible mechanism for inducing the appearance of non-zero classical fields
$n_{\mu }$ and $n_{\mu }^{i}$, as the VEVs of the original vector fields
$A_{\mu }$ (\ref{f}) and $A_{\mu }^{i}$ (\ref{ab}).
The simplest case to consider involves just
vector and scalar field interactions.

In the Abelian case, with one vector and one real scalar field, the
corresponding general Lagrangian density is given by:
\begin{equation}
L=-\frac{1}{4}F_{\mu \nu }F_{\mu \nu }+\frac{h}{2}A_{\mu }^{2}\sigma ^{2}+
\frac{1}{2}(\partial _{\mu }\sigma )^{2}-\frac{1}{2}\mu ^{2}\sigma ^{2}-
\frac{\lambda }{4}\sigma ^{4}+L^{\prime}  \label{sig}
\end{equation}
Here we have divided the total Lagrangian into two parts: the explicitly
written terms are, as we will see, compatible with Lorentz invariance, while
the part $L^{\prime}$ includes all other terms having an independent
structure, which break Lorentz invariance once the SBLS appears. These
include a fundamental mass term for the vector field
$\frac{M^{2}}{2}A_{\mu}^{2}$ and a quadrilinear term
$A_{\mu }^{2}\cdot A_{\mu }^{2}$, both
of which were considered in section \ref{abelian}. Also there are the
interaction terms $\partial _{\mu }A_{\nu }\cdot A_{\mu }A_{\nu }$,
$A_{\mu}\cdot \sigma \partial _{\mu }\sigma $
and all the odd-power couplings of
the $\sigma $ (which for simplicity we exclude, by requiring the reflection
$Z_{2}$ symmetry $\sigma \rightarrow -\sigma $ ).

Let us consider first the possible vacuum configurations corresponding to
the potential in $L$ (\ref{sig}):
\begin{equation}
U=-\frac{h}{2}A_{\mu }^{2}\sigma ^{2}+\frac{1}{2}\mu ^{2}\sigma ^{2}+
\frac{\lambda }{4}\sigma ^{4}+U^{\prime }  \label{potential}
\end{equation}
One can see that the explicitly written part of the
potential $U$ (\ref{potential}) drives the vector
field $A_{\mu }$ to have a VEV as large as
possible---it is in fact unstable in the absence of a quadrilinear term
$A_{\mu }^{2}\cdot A_{\mu }^{2}$, or some other stabilizing term
(see below). For the moment, let us ignore
$U^{\prime }$ and this instability and take the VEV of $A_{\mu }$ as some
given parameter. Variation of the potential $U$ with respect to $\sigma $
yields one equation between the VEVs $n_{\mu }$ and $v$ of the fields (for
$v\neq 0$):

\begin{equation}
A_{\mu }=a_{\mu }+n_{\mu },\quad \sigma =\rho +v,\textrm{ \quad }hn_{\mu
}^{2}=\lambda v^{2}+\mu ^{2}\textrm{ }  \label{vevs}
\end{equation}
This leads to a non-trivial minimum of the potential $U$

\begin{equation}
U_{\min }=-\frac{1}{4\lambda }(hn_{\mu }^{2}-\mu ^{2})^{2}  \label{min}
\end{equation}
provided that $hn_{\mu }^{2}>$ $\mu ^{2}$. The trivial minimum $U_{\min }=0$
with $v=0$ corresponds to the case $hn_{\mu }^{2}\leq \mu ^{2}$. As
mentioned above, the potential $U$ is unbounded below and drives the VEV of
$A_{\mu }$ towards large values, $hn_{\mu }^{2}\rightarrow \infty $.
So it is necessary to add a term to the potential $U$ in order to cut-off
this instability for some large (but finite) value of
$A_{\mu}^2$. As a simple example, we could add a cut-off
term $U_{cut-off} = (A_{\mu}^2)^2\theta\left((A_{\mu}^2)^2
- \Lambda^4 \right)$
and consider the limit $\Lambda^2 \rightarrow \infty$ at the end
of the calculation, when Lorentz
symmetry breaking effects from such a term become unobservable.
Similar results can be obtained by including a large
dimensional irrelevant term, such as
$U_{cut-off} =\frac{1}{6}\frac{(A_{\mu}^2)^3}{\Lambda^2}$;
however note that it is necessary to take $\mu^2$ (and $v^2$)
to be of order the cut-off scale $\Lambda^2$.

Now, apart from the standard symmetrical case ($n_{\mu }=0$, $v=0$), the
following vacua are possible for the potential $U$ (depending
on the sign and value of $n_{\mu }^{2}$
and $\mu ^{2}$): an ordinary Goldstone-Higgs (GH) mode ($n_{\mu }=0$, $v\neq
0$) which breaks the above reflection symmetry, the SBLS mode ($n_{\mu }\neq
0$, $v=0$) and the combined SBLS-GH mode ($n_{\mu }\neq 0$, $v\neq 0$).
When the cut-off term $U_{cut-off}$ is added, one finds that
the preferred vacuum configuration in the potential $U$
corresponds to the case when the vector field $A_{\mu }$
develops a VEV $n_{\mu }$, having a large
value $|n_{\mu }^{2}| = \Lambda^2$.
The pure GH minimum is degenerate with the symmetrical one
($U_{\min }=0$) for $\mu ^{2}>0$ or located at
$U_{\min} = -\frac{\mu^4}{4\lambda}$ for $\mu ^{2}<0$.
This means that, provided $h\Lambda^{2}>  |\mu ^{2}| $,
the combined SBLS-GH mode dominates over the other
possible vacua\footnote{The pure SBLS mode
($n_{\mu }\neq 0$, $v=0$) corresponds to the particular
choice $n_{\mu }^{2}=\frac{\mu ^{2}}{h}$ and is degenerate with the trivial
symmetrical case with $U_{\min }=0$.}.

After symmetry breaking (\ref{vevs}), the potential $U$ (\ref{potential})
takes the form

\begin{equation}
U=-\frac{h}{2}(a_{\mu }+n_{\mu })^{2}(\rho +v)^{2}+\frac{1}{2}\mu ^{2}(\rho
+v)^{2}+\frac{\lambda }{4}(\rho +v)^{4}+U^{\prime}(a_{\mu}+n_{\mu},\rho+v)
\label{poten}
\end{equation}
It can be seen from (\ref{poten}) that, due to the SBLS and the spontaneous
breakdown of the reflection $Z_{2}$ symmetry\footnote{Another physical
consequence is the formation of domain walls separating the
vacua with $<\sigma >=+v$ and $<\sigma >=-v$.} $\sigma \rightarrow -\sigma $,
the whole vector field $A_{\mu }=a_{\mu }+n_{\mu }$ has acquired a soft
mass in addition to the fundamental mass term contained in $U^{\prime}$.
However, the situation drastically changes when $U^{\prime}$, which
simultaneously induces (as we already know from section \ref{abelian}) the
physical breakdown of Lorentz symmetry, is required to vanish identically.
Thereby, the potential (\ref{poten}) becomes nothing but the potential of
spontaneously broken scalar electrodynamics. This can be converted into its
well-known gauge invariant form by making
an inverse Higgs transformation:

\begin{equation}
a_{\mu }\rightarrow a_{\mu }+\frac{1}{gv}\partial _{\mu }\theta \textrm{ ,
\quad }\phi =\frac{\rho +v}{\sqrt{2}}e^{i\frac{\theta }{v}}\textrm{ , \quad }
\phi ^{*}=\frac{\rho +v}{\sqrt{2}}e^{-i\frac{\theta }{v}}  \label{gauge}
\end{equation}
So, one finally arrives at the standard QED Lagrangian

\begin{eqnarray}
L &=&-\frac{1}{4}F_{\mu \nu }F_{\mu \nu }+ig(a_{\mu }+n_{\mu })\left( \phi
^{\star }\partial _{\mu }\phi -\partial _{\mu }\phi ^{\star }\phi \right) +
\label{qed} \\
&&+g^{2}(a_{\mu }+n_{\mu })^{2}\phi ^{\star }\phi +\partial _{\mu }\phi
^{\star }\partial _{\mu }\phi -\mu ^{2}\phi ^{\star }\phi -\lambda \left(
\phi ^{\star }\phi \right) ^{2}\textrm{.}  \nonumber
\end{eqnarray}
of the massive charged scalars $\phi $ and $\phi ^{*}$ with a gauge coupling
constant $g=\sqrt{h}$. Going then by a unitary transformation to a new
scalar field $\phi $ $\rightarrow e^{ign\cdot x}\phi $, one can completely
eliminate\footnote{The consequent spontaneous breakdown of momentum
conservation, mentioned in footnote \ref{foot1} and
postulated to be unobservable since it also violates
Lorentz symmetry, is in fact hidden by the global charge conservation
manifest in the Lagrangian (\ref{qed}). This global charge conservation
arose through the inverse Higgs transformation (\ref{gauge}).}
any dependence on the constant vector $n_{\mu }$ in the Lagrangian
(\ref{qed}) thus excluding it from the VEV of the charged field $\phi $
(\ref{gauge}) as well. This means that the predominant
SBLS-GH mode, while being
physical in the Lorentz non-invariant phase, turns into a false vacuum when
Lorentz invariance is restored and gauge symmetry appears. Thus, any terms
in the original Lagrangian (\ref{sig}) which have not evolved to a gauge
invariant form (they are collected in $L^{\prime }$) must definitely
vanish---otherwise the SBLS-GH mode will dominate, inevitably inducing an
explicit Lorentz non-invariance. The actual scenario for the physical vacuum
depends only on the mass of the scalar field $\phi $. If $\mu ^{2}$ is
positive one has standard scalar electrodynamics with a massless photon, if
it is negative then a spontaneously broken gauge theory emerges. So, we come
to the conclusion that spontaneously broken gauge symmetries could also
appear when the SBLS happens, as a consequence of the non-observability of
the Lorentz symmetry breakdown.

The emergence of the massive charged scalar $\phi $ (or neutral scalar $\rho
$ in the case of the spontaneously broken gauge theory) is, as a matter of
fact, the most direct prediction of the model considered. In the non-Abelian
case, to which this model is straightforwardly extended, it follows from the
generalisation of Eq.~(\ref{gauge}) that the scalars should belong (like the
vector fields do) to the adjoint representation of the gauge symmetry group.

In the model of section \ref{abelian} with a charged fermion
field, the SBLS arose as a consequence of the existence
of ``flat directions'', meaning
that from gauge symmetry the energy for a vacuum state with $A_{\mu} = 0$
is the same as for one with $A_{\mu}= n_{\mu}$. If the reader does
not like this idea of getting SBLS as the result of a flat direction,
the price to be paid is all the above scalars.
Alternatively, instead of using fundamental vector fields $A_{\mu}^i$,
one could consider them as being effective
composite fields and then again ask the question
from which we started in this section: what could induce the SBLS?
Some possibilities may
be related, for example, with a Lorentz-symmetry breaking
fermion-antifermion pair condensate, as in the composite models mentioned
earlier \cite{bj}, rather than with
the scalar condensate considered here. However,
independent of its origin, if SBLS appears and is unobservable, it
inevitably generates some gauge invariance, as was demonstrated in the
previous sections.

\section{Conclusion \label{conclusion}}

We have shown that gauge invariant Abelian and non-Abelian theories can be
obtained from the requirement of the physical non-observability of the SBLS
rather than by using the Yang-Mills gauge principle.

Imposing the condition that the Lorentz symmetry breaking be unobservable of
course restricts the values of the coupling constants and mass parameters in
the Lagrangian density. These restrictions may naturally also depend on the
direction and strength of the Lorentz symmetry breaking vector field vacuum
expectation values (VEVs), whose effects are to be hidden. This allows us a
choice as to how strong an assumption we make about the non-observability
requirement. Actually, in the Abelian case, we just assumed this
non-observability for the physical vacuum
that really appears (\ref{f}), but we needed a stronger assumption in the
non-Abelian case.
A stronger assumption is needed because
we believe that, for a given translational
invariant but Lorentz symmetry breaking vacuum state, it is always possible
to hide SBLS by only imposing restrictions on the couplings which lead
to an Abelian gauge theory.
We could ensure the derivation of a non-Abelian gauge symmetry by
choosing a stronger version of the non-observability assumption:
the SBLS is unobservable for all conceivable vacuum field
configurations, including even those vacua that could never be
realised. This is clearly a very strong assumption. Then we note that we
let the coupling constants and mass parameters be restricted in a way which
a priori depends on the vacuum from which we have to hide the SBLS. So, as
the vacuum state itself depends on the values of these coupling constants
and mass parameters, there is a back reaction problem unless one simply
imposes that all possible vacua must have their SBLS hidden. Nevertheless
this assumption looks to be too strong, in the sense that even Yang-Mills
theories would in general not be able to hide the SBLS unless
all the $n_{\mu}^i$ commute. The reason for this is that, if the
$n_{\mu}^i$ did not commute, the Yang-Mills field tensor
$F_{\mu\nu}^a$ would not vanish in the vacuum
and would break Lorentz symmetry spontaneously. If,
for some combination of the Lorentz indices $\mu$, $\nu$ we
have some components $F_{\mu\nu}^a \ne 0$,
there is no way to gauge that property away.
Thus once the field tensor $F_{\mu\nu}^a$ is non-zero, it is virtually
impossible to gauge the SBLS away. We have therefore invented a milder
assumption, which can be used for deriving non-Abelian
gauge symmetry: the SBLS is unobservable in any vacuum for which
the vector fields have VEVs  of the factorised form
$n_{\mu}^i = n_{\mu} \cdot f^i$ (\ref{fff}). This factorised form
is a special case in which the $n_{\mu}^i$ commute with each other.

A working assumption in the present work is that the vector
fields are taken to be pure spin-1 fields, satisfying the Lorentz
condition (\ref{t}). Due to this constraint
$\partial_{\mu}A_{\mu} = 0$, there is a difficulty, a priori, in
prescribing the functional measure ${\cal{D}}A_{\mu}$
over the vector fields in
the path integral for the theory. Therefore a priori we have only
worked in the classical approximation (although most of our arguments
are also true quantum mechanically). In the Abelian case it is
essentially obvious how to define the measure ${\cal{D}}A_{\mu}$.
However, in the non-Abelian case, it would be necessary to
introduce ghost fields as usual, in order to keep the gauge
invariance beyond tree-level. We do not address here the interesting
and important question of how these ghost fields arise in our
argumentation, when calculating quantum mechanically.
At the same time, we recall from footnote 3 that the Lorentz
condition (\ref{t}) is not actually required, when we use the strong
form of the SBLS non-observability assumption and space-time
dependent vacuum vector fields.

Gauge transformations having parameters $\omega^i(x) = n^i \cdot x$
varying linearly with $x^{\mu}$
are important in our calculations, in as far as they induce changes in
the VEVs of the vector fields. Such linear gauge functions
$\omega^i(x)$ do not
obey the boundary conditions of vanishing at infinity---not even
their gradients vanish at infinity. However, due to the required
commutation of the different components $n_{\mu}^i$ discussed above,
we effectively work with a Cartan algebra. Since we are only using
a Cartan algebra, the immediately needed large fields at infinity do not
imply the large gauge transformations required to allow
non-Abelian instantons (it is essential that the instantons are genuinely
non-Abelian even at infinity). So the danger of introducing
non-trivial topological effects is avoided.

There could of course be other (non-perturbative) sources of
spontaneous breaking of Lorentz invariance, causing the field
strengths $F_{\mu\nu}^a$ to acquire non-zero VEVs in Abelian
or non-Abelian models. In this case, however, there would be no way
to rescue the Lorentz invariance, even by having a gauge symmetry
in the theory. Such a theory would be totally excluded by our
main assumption of no observable Lorentz invariance breaking.
So logically our strong requirement of the non-observability
of SBLS excludes this possibility, which otherwise could easily
happen. It was precisely to avoid this possibility that we
introduced our assumption of factorised VEVs (\ref{fff}) in
the non-Abelian case.

Remarkably, the application of the here proposed non-observability principle
to the most general relativistically invariant Lagrangian density, with
arbitrary couplings for all the fields involved, leads by itself to the
appearance of a continuous symmetry in terms of conserved
Noether currents and, what is more, to the massless vector field(s) gauging
this symmetry. As a consequence, matrices constructed out of the coupling
constants constitute representations of the Lie algebra corresponding to the
above symmetry and (in the non-Abelian case) the coupling constants for the
vector field self-interaction terms prove to be the structure constants for
this algebra. Thus the vector fields are then a source of the symmetries,
rather than local symmetries being a source of the vector fields as in the
usual formulation\footnote{In a similar manner,
it was shown a long time ago \cite{ogievetsky} that the
massive Abelian or non-Abelian theories can be obtained from the requirement
that the vector fields preserve their transversality (\ref{t}) in all their
interactions. However, this requirement cannot lead to the massless vector
field case, and for good reason.}. We considered a simple model for SBLS
based on the interaction between a massive vector and a real scalar
field and found that the gauge, or spontaneously broken gauge, symmetry
phase has to appear. So, to conclude, when Lorentz symmetry spontaneously
breaks through the appearance of constant background vector fields then,
simultaneously, a stable gauge (or spontaneously broken gauge) symmetry
phase is created in order to avoid a physical breakdown of Lorentz
invariance.

It is difficult to confirm our SBLS hypothesis phenomenologically since,
by assumption, there can be no non-Lorentz invariant evidence for it.
However, the very fact that there are many
well-established gauge symmetries in the Standard Model might be taken as
a weak confirmation of the viewpoint in this paper.
There could be some indirect phenomenological manifestation of
this picture, related to the underlying mechanism responsible for
the genesis of the SBLS, such as the existence of scalar or
composite fermion states at higher scales. We leave this and
other open questions for further investigation.

\section*{Acknowledgements}

We are indebted to David Sutherland for many suggestions and useful remarks.
One of us (JLC ) would like to thank Oleg Kancheli for stimulating
conversations and criticism and Zurab Berezhiani, Hans Durr, Kazumi Maki
and Peter Minkowski for interesting discussions.
Financial support to JLC from INTAS grant No.
96-155, PPARC grant No. PPA/V/S/2000/ and the Joint Project grant from the
Royal Society are also gratefully acknowledged. CDF and HBN would like to
thank the EU commission for financial support from grants No. SCI-0430-C
(PSTS) and No. CHRX-CT-94-0621. Also JLC, CDF and HBN would like to
acknowledge financial support from INTAS grant No. 95-567.

\end{document}